\documentclass[11pt,twoside]{article}
\usepackage{asp2004}
\usepackage{graphicx}
\usepackage{lscape}

\begin{document}

\markboth{Schreiber, Hameury, and Lasota}{Predictions of the disc
instability model} \pagestyle{myheadings}

\title{Predictions of the disc instability model}
\author{M.R. Schreiber}
\affil{Obs. de Strasbourg, 11 rue
de l'Universit\'e, 67000 Strasbourg, France, \\
AIP, An der Sternwarte 16, D-14482 Potsdam, Germany}
\author{J.-M. Hameury}
\affil{UMR 7550 du CNRS, Observatoire de Strasbourg, 11 rue de
l'Universit\'e, F-67000 Strasbourg, France}
\author{J.-P. Lasota}
\affil{Institut d'Astrophysique de Paris, 98bis Boulevard Arago,
75014 Paris, France}

\begin{abstract}
Confronting the predictions of numerical calculations with observations is an
important tool to progress with our understanding of the mechanism triggering
the outbursts of dwarf novae and soft X-ray transients. Simultaneous
multi-wavelength observations are of particular importance in this process as
they contain information about the chronology of the outbursts as e.g. the
famous UV-delay observed in dwarf novae. We review key-predictions of
the disc instability model (DIM) and confront them  with the corresponding
observations.
\end{abstract}

\section{Introduction}
Dwarf novae  are a subgroup of cataclysmic variables (CVs) whose
rather regular $2-5$\,mag outbursts are thought to be driven by
disc instabilities. This idea is based on the existence of a
thermal-viscous instability in regions where hydrogen is partially
ionized, and opacities depend strongly on temperature. Disc
instabilities as the mechanism for dwarf nova outbursts have been
proposed almost 30 years ago and first numerical models have been
developed in the eighties. Compared to these early models the
current disc instability model (DIM) has become rather complex:
evaporation \citep{meyer+meyer-hofmeister94-1}, irradiation
\citep{hameuryetal99-1,schreiber+gaensicke01-1}, stream impact
heating, tidal dissipation \citep{buat-menardetal01-1}, stream
overflow \citep{schreiber+hessman98-1}, magnetic truncation
\citep{lasotaetal95-1}, and mass transfer variations on almost
every possible time scale
\citep{schreiberetal00-1,buat-menardetal01-2} have been added to
the ``pure'' disc instability. Consequently, the number of
unconstrained parameter significantly increased. Confronting
predictions of the model with observations requires therefore to
focus on systematic dependencies of the predictions on the
parameter instead of trying to ``fit'' the observations.

We review predictions of state-of-the-art DIM calculations in the
light of recent observations. In particular, we concentrate on
four topics: UV and EUV delays  (Sect.\,2); truncation and X-ray
emission during quiescence (Sect.\,3); the triggering mechanism
for superoutbursts in SU~UMa systems (Sect.\,4); and the absolute
magnitude of SS\,Cyg

\section{Delays: outside-in versus inside-out outbursts}
Since the early days of the model, it has been known that
outbursts can be triggered close to the white dwarf or in outer
disc regions. If the mass transfer rate is high, the accumulation
time at the outer disc edge can be shorter than the viscous
diffusion time, and the instability will be triggered in the disc
outer regions; the outburst is of the outside-in type. On the
other hand, for low mass transfer rates, the viscous time is 
shorter, and the outburst will be triggered at the inner edge;
it is of the inside-out type. The limit between both types
of outburst therefore depends sensitively on parameters such as
the mass transfer rate and the viscosity; it is therefore
important to be able to determine the type of observed outbursts
for constraining these parameters.

The so-called UV delay between the UV and optical rise, measured
for several dwarf novae, has often been considered as a power
discriminating tool between both types of outbursts. As UV
radiation is emitted close to the white dwarf, one would expect
that there should be a long delay in cases where the outburst is
triggered in the disc outer regions, and no delay when the
outburst starts at the inner disc edge. That this hypothesis can
explain the observations has been stated early by
\citet{smak84-1}. Shortly thereafter \citet{pringleetal86-1}
claimed the existence of a ``problem of the UV-delay'' with the
consequence that the following years were characterized by many
attempts to find a solution for the alleged problem. Finally
\citet{smak98-1} came to the conclusion, that there is no
problem if one uses the correct outer boundary condition. Here we
show that Smak is right in that there is no problem but in
contrast to his findings this has nothing to do with the outburst
type.

Figure 1 shows predicted optical, UV, EUV, and X-ray light curves
assuming the orbital parameter of SS\,Cyg. These light
curves are calculated using the DIM-code described in detail in
\citet[][]{hameuryetal98-1} in which tidal dissipation and stream
impact heating have been included \citep{buat-menardetal01-1} \footnote{
Please note, our general conclusions are independent of this effects}. 
The emission from the boundary layer is approximated as described in
\citet{schreiberetal03-1}: the boundary layer is optically thin
and emits X-rays for low accretion rates, but becomes optically
thick when the accretion rate exceeds the critical value
$\dot{M}_{\rm cr}=10^{16}$\,gs$^{-1}$.
\begin{figure}[t]
\center
\includegraphics[width=7.8cm, angle=270]{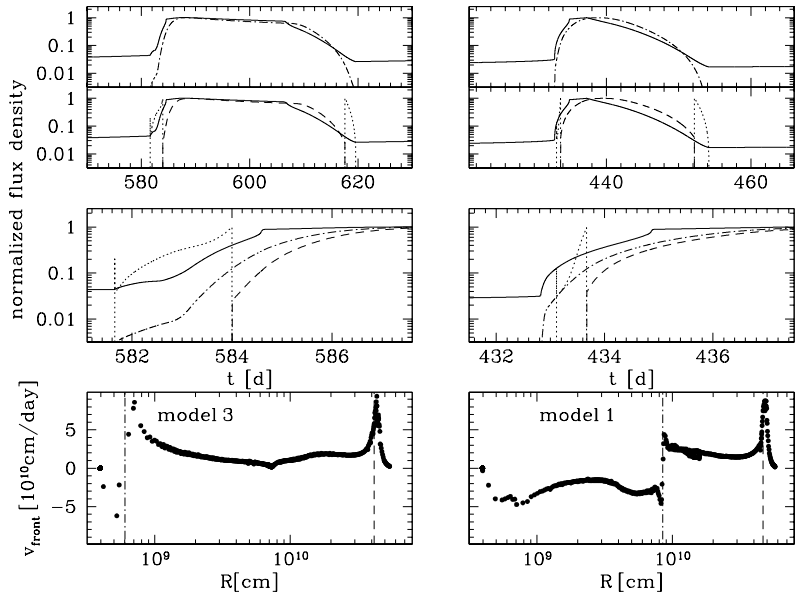}
\caption{The predicted normalized optical (solid line) and UV flux
(dashed--dotted) as well as the normalized boundary layer
emission, i.e. X-rays (dotted) and EUV (long dashed) for
inside-out (left) and outside-in (right) outbursts. The
intermediate panel shows a detailed view of the outburst rise. The
bottom panels show the heating front velocity corresponding to the
rise of the outbursts. The ignition radius is marked by the
vertical dashed--dotted line \citep[see
also][]{schreiberetal03-1}.}
\end{figure}

Apparently, the calculated UV and EUV delays are {\em longer} in
the case of an inside-out outburst - just the opposite to what one
would naively expect. To understand this result one has to take
into account that the UV emission is rising when inner disc
regions reach high accretion rates which is {\em not} identical to
the moment at which the heating front arrives. The same holds for
the EUV rise: to get a significant increase of EUV emission, the
accretion rate onto the white dwarf has to increase and this does
{\em not} necessarily happen when the heating front is reaching
the boundary layer.

One has to consider another difference of outside-in and
inside-out heating fronts than their propagation direction to
understand the light curves. Inside-out heating fronts must
propagate ``uphill'' against the surface-density and
angular-momentum gradients. This makes the propagation difficult
and such fronts can be subject to dying before reaching the outer
disc \citep[see, e.g.][]{lasota01-1}. In contrast, an
outside--in heating front starts in high surface density regions
and has an easy way ``downhill'' sliding down along the gradients.
They never die before fulfilling their task. Thus, for an
outside--in front the accretion rate in the inner regions of the
disc as well as the accretion rate onto the white dwarf increase
much more rapidly than in the case of an inside--out heating
front. As a consequence, the UV and EUV delay at the onset of the
optical rise are {\em shorter} than in the inside--out case but
both are in agreement with the observations
\citep{schreiberetal03-1}. The old picture is valid only for the
X-ray emission from the optically thin boundary layer: in the case
of an inside-out outbursts there is no delay of the X-ray rise
with respect to the optical (Figure 1). When the heating
front started far away from the white dwarf, the X-ray flare is
delayed relative to the optical rise by $0.3-0.45$\,days, i.e. the
time it takes the heating front to reach the inner edge.

\section{X-rays and truncation during quiescence}

The accretion rate derived from X-ray observations of dwarf novae
during quiescence is generally several orders of magnitudes larger
than the one predicted by the standard DIM. Three prominent
examples are SS\,Cyg \citep{wheatleyetal03-1}, VW\,Hyi
\citep{pandeletal03-1}, and OY\,Car \citep{ramsayetal01-2}. As it
is too often ignored we want to stress here again that disc
truncation by a magnetic field
\citep{lasotaetal95-1,hameuryetal97-1,schreiberetal03-1,schreiberetal04-1}
and/or by evaporation \citep{meyer+meyer-hofmeister94-1} can bring
into agreement models and observations because {\em{\bf the disc
is not stationary during quiescence}}.
\begin{figure}[t]
\center
\includegraphics[width=6.5cm, angle=270]{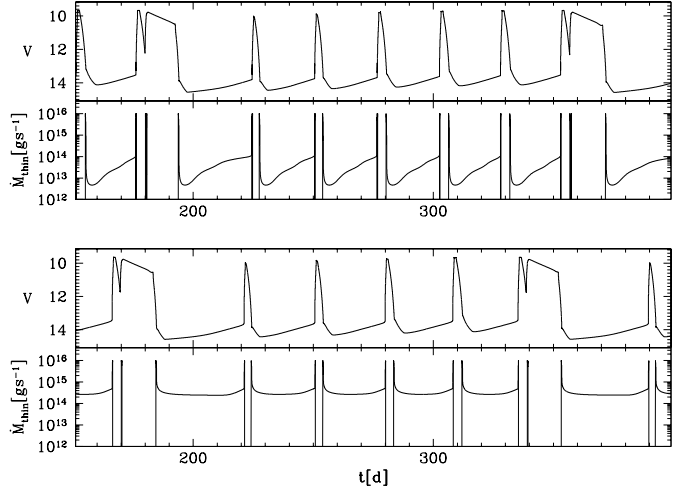}
\caption{\label{f-xevol} Optical light curves and mass accretion
rates when the inner region is assumed to emit X-rays
($\dot{M}_{\rm thin}$) assuming the orbital parameters of VW\,Hyi.
In the top panels we assumed the disc to extend
down to the surface of the white dwarf whereas the disc has been
truncated during quiescence in the bottom panels. In the latter
case the accretion rate is higher and almost constant during
quiescence. X-ray flares are predicted at the onset and end of the
outbursts.}
\end{figure}
Assuming e.g. a weakly magnetic white dwarf, the inner disc radius
is given by the ``magnetospheric'' radius, i.e. $R_{\mathrm{in}}=
R_{\mathrm{M}}= 9.8\times10^8 \dot{M}_{15}^{-2/7}M_{\rm
wd}^{-1/7}\mu_{30}^{4/7}\mathrm{cm}$ where $\mu_{30}$ is the
magnetic moment of the white dwarf in units of
$10^{30}$\,G\,cm$^3$ \citep[][]{hameury+lasota02-1}. Th
postulation of such an inner hole dramatically increases the
predicted X-ray flux during quiescence. Figure 2 displays this
influence of truncation on the predicted X-ray emission for the
parameter of VW\,Hyi. 
Instead of $10^{11}-10^{13}$\,g\,s$^{-1}$ without truncation we obtain
$\dot{M}_{\rm acc} \sim\,3-5\times10^{14}$\,g\,s$^{-1}$ with
truncation. This increase of the expected X-ray emission during
quiescence results from the fact that the DIM predicts accretion
rates which increase with radius while the disc accumulates mass.
\section{What triggers superoutbursts in SU~UMa systems?}
SU~UMa stars are short-period, i.e. $P_{\rm orb} \leq\,2.2$\,hr
dwarf novae whose light curve consists of two types of outburst:
normal dwarf nova outbursts and 5-10 times longer as well as
$\sim0.7$\,mag brighter superoutbursts. Sometimes superoutbursts
follow a so-called precursor outburst, i.e. a normal outburst
shortly before the superoutburst.

Since the early days of the DIM it is discussed whether the
particular behaviour of SU~UMa systems is caused by a thermal
tidal instability (TTI) \citep[e.g.][]{osaki96-1} or by enhanced
mass transfer (EMT) from the secondary \citep{hameuryetal00-1}.
Here we focus just on one important systematic difference of the
models which concerns the sensitivity of the predictions to
variations of the mean mass transfer rate. In the TTI model (TTIM)
a superoutburst is triggered when the disc expands beyond the 3:1
resonance radius and becomes tidally unstable. Increasing the mass
transfer rate has two effects: (1) the disc's mass increases
faster which makes it easier to reach the critical radius during
outburst and (2) the torque exerted by the stream material
increases which makes it more difficult to reach
$R_{\mathrm{crit}}$. Both effects cancel out to some extent and
the predictions of the TTIM are rather insensitive to fluctuations
of the mass transfer rate. In the EMT model (EMTM) the triggering
condition for superoutbursts is independent of the outer radius
reached during the expansion. Instead it depends on the maximum
accretion rate during outburst which increases with the mass of
the disc. So the disc mass must exceed a critical value to trigger
a superoutburst. The time scale on which a superoutburst is
triggered depends therefore {\em only} on the mean mass
transfer rate.

Figure 3  shows light curves calculated with the TTIM and the EMTM
for different values of the mass transfer rate. While the mass
transfer rate has been changed drastically in the case of the TTIM
(Figure 3, right), we just slightly increased it for the EMTM
(Figure 3, left). The predictions of the EMTM are much more
sensitive to mass transfer variations as e.g. the number of normal
outbursts decreased from five to three and the precursor outburst
disappeared. The burning question now is, how does this difference
relate to observations?
\begin{figure}\begin{center}
\includegraphics[width=.44\textwidth, angle=0]{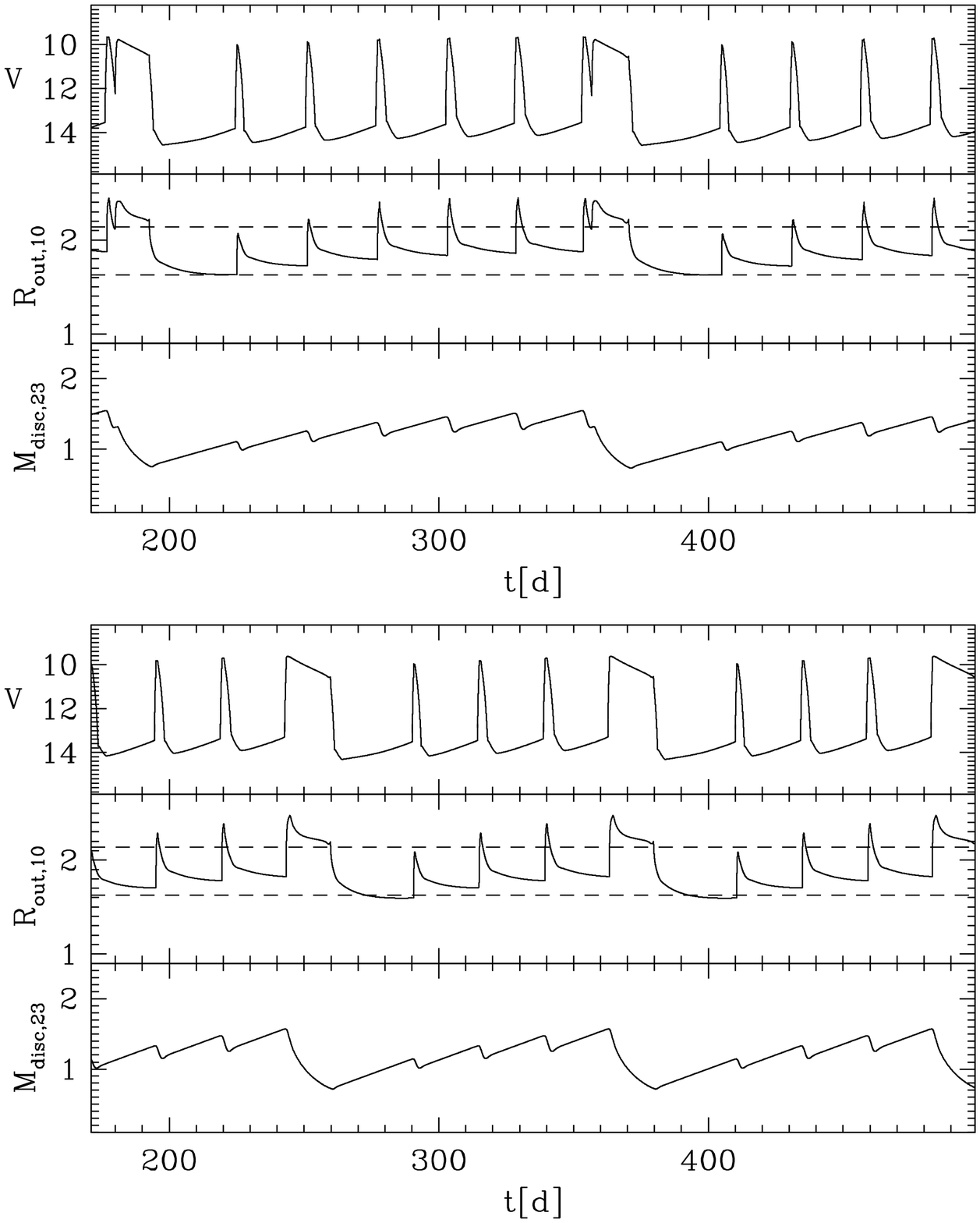}
\includegraphics[width=.44\textwidth, angle=0]{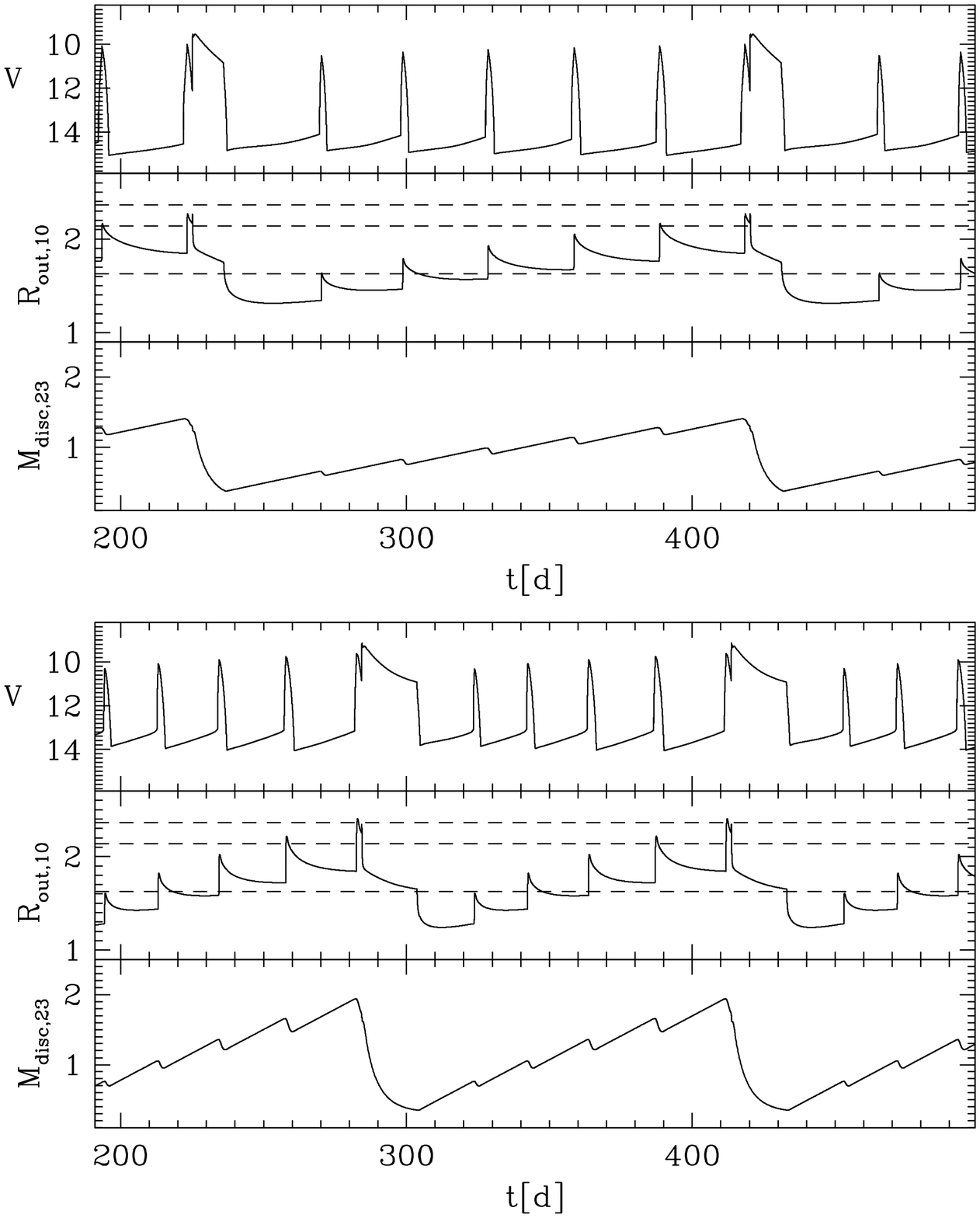}
\end{center}
\caption{The response of the TTIM (right) and the EMTM (left) to
strong
  (right) and small (left) changes of the mean mass transfer rate. Concerning
  the outburst cycle, the number of normal outbursts per supercycle as well as
  the precursor phenomenon, the predictions of the EMTM depend sensitively
  depend on the mean mass transfer rate \citep[for more details see
  also][]{schreiberetal04-1}. }
\end{figure}

The answer is obvious. Observed light curves of SU~UMa stars
do not only differ from system to system, there are also strong
variations in the light curves of single systems. For example, in
VW\,Hyi neither the single outbursts nor the outburst
cycles are strictly periodic. \citet{bateson77-1} divided the
superoutbursts of VW\,Hyi in two classes: those with a
single superoutburst and those where a precursor outburst is
separated from the superoutburst. In addition, the number of
normal outbursts observed between two superoutbursts varies from
three to seven and the supercycle duration ranges from $\sim\,100$
to $\sim\,250$ days \citep[e.g.][]{mohanty+schlegel95-1}. Such
variations of the supercycle length and the frequency of normal
outbursts can be considered as typical for ordinary SU~UMa stars.
They are natural outcome of the EMTM while the TTIM fails to
explain them.

\section{The DIM and the distance of SS\,Cyg}

The HST/FGS parallax of SS\,Cyg revealed a distance of
$166\pm12$\,pc \citep{harrisonetal99-1} and it has been shown that the
required absolute magnitude during outburst cannot be reproduced  by the
current DIM \citep{schreiber+gaensicke02-1}. If the extreme ``new''
distance were confirmed, two distinct problems would appear. First, one
would have to explain why SS\,Cyg has behaved for more than 100 years as
a bona fide U\,Gem dwarf nova, but is on the average brighter than Z\,Cam 
and two other DN of the same type
EM Cyg and AH Her, which are at the
same orbital period, and presumably have a larger disc due to a less
massive primary. Second, even if this observational problem could be
solved by e.g. reevaluating distances to {\em all} CVs, one would still
have to (at least) drastically revise the DIM as the critical
temperature above which the disc is stable is one of the few predictions
of the model which does not depend on any free parameter; this would
imply decoupling the change of viscosity from the ``original''
instability due to partial ionization of hydrogen -- a frightening
hypothesis as the existence of the ``natural'' instability is one of the
key-strengths of the model.

\bibliographystyle{aa}


\end{document}